\documentclass[9pt,twocolumn,twoside]{osajnl}
\journal{josaa}
\setboolean{shortarticle}{false} % true = letter, false = research article
\usepackage{amsmath,amsfonts,amssymb}
\usepackage{graphicx}
\usepackage{wrapfig}
\usepackage{color,soul}
\long\def\comment#1{}

\def\bA{\boldsymbol{A}}
\def\ba{\boldsymbol{a}}
\def\ra{\mathrm{a}}

\def\bc{{\bf c}}

\def\bC{\boldsymbol{C}}

\def\bD{\boldsymbol{D}}
\def\bd{\boldsymbol{d}}

\def\bE{\boldsymbol{E}}
\def\rE{\mathrm{E}}

\def\bg{{\bf g}}
\def\bg{\boldsymbol{g}}
\def\bG{\boldsymbol{G}}

\def\bH{\boldsymbol{H}}

\def\bi{\boldsymbol{i}}

\def\ri{\mathrm{i}}

\def\rm{\mathrm{m}}

\def\bM{\boldsymbol{M}}
\def\bn{\boldsymbol{n}}
\def\rn{\mathrm{n}}

\def\bp{\boldsymbol{p}}
\def\rp{\mathrm{p}}
\def\bP{\boldsymbol{P}}

\def\bQ{\boldsymbol{Q}}
\def\br{\boldsymbol{r}}

\def\bR{\boldsymbol{R}}

\def\bS{\boldsymbol{S}}

\def\bT{\boldsymbol{T}}
\def\rT{\mathrm{T}}

\def\bu{\boldsymbol{u}}
\def\bv{\boldsymbol{v}}

\def\bw{\boldsymbol{w}}

\def\bx{\boldsymbol{x}}
\def\by{\boldsymbol{y}}

\def\bZ{\boldsymbol{Z}}
\def\balpha{\boldsymbol{\alpha}}

\def\bchi{\boldsymbol{\chi}}
\def\bmu{\boldsymbol{\mu}}

\def\bXi{\boldsymbol{\Xi}}
\def\bnu{\boldsymbol{\nu}}

\def\bSigma{{\bf \Sigma}}

\title{Millisecond Exoplanet Imaging, II: Regression Equations and Technical Discussion}

\author[1,*]{Richard A. Frazin}
\author[2]{Alexander T. Rodack}
\affil[1]{ Dept. of Climate and Space Sciences and Engineering, University of Michigan, Ann Arbor, MI 48109}
\affil[2]{Steward Observatory, University of Arizona, Tucson, AZ 85721}
\affil[*]{E-mail: rfrazin@umich.edu}

\dates{Compiled \today}

\ociscodes{010.1080 Adaptive Optics, 010.7350   Wavefront Sensing}

\doi{\url{http://dx.doi.org/10.1364/ao.XX.XXXXXX}}

\begin{abstract}

The leading difficulty in achieving the contrast necessary to directly image exoplanets and associated structures (e.g., protoplanetary disks) at wavelengths ranging from the visible to the infrared are quasi-static speckles, and they are hard to distinguish from planets at the necessary level of precision.
The source of the quasi-static speckles is hardware aberrations that are not compensated by the adaptive optics (AO) system.
These aberrations are called non-common path aberrations (NCPA) by the community.
In 2013, Frazin showed how, in principle, simultaneous millisecond (ms) telemetry from the wavefront sensor (WFS) and the science camera behind a stellar coronagraph can be used as input into a regression scheme that simultaneously and self-consistently estimates the NCPA and the sought-after image of the planetary system (the \emph{exoplanet image}).
The physical principle underlying the regression method is rather simple: the wavefronts, which are measured by the WFS, modulate the speckles caused by the NCPA and therefore can be used as probes of the optical system.
The simulations in the Part I article provide results on the joint regression on the NCPA and the exoplanet image from three different methods, called the \emph{ideal}, the \emph{na\"ive}, and the \emph{bias-corrected} estimators.
The ideal estimator is not physically realizable but is a useful as a benchmark for simulation studies, but the other two are, at least in principle.
This article provides the regression equations for all three of these estimators as well as a supporting technical discussion.
Briefly, the na\"ive estimator simply uses the noisy WFS measurements without any attempt to account for the errors, and the bias-corrected estimator uses statistical knowledge of the wavefronts to treat errors in the WFS measurements.
\end{abstract}

\begin{document}
\maketitle

\section{Introduction}

This article is the second in series on the use of millisecond exposures for ground-based direct imaging of exoplanets using a stellar coronagraph.   
We will refer to the first article as \emph{Part I}.\cite{PartI}
The context, motivation, references to the literature and related introductory material are provided in the introduction of Part I, and it should be considered as prerequisite to reading this article.

\begin{figure}
	\includegraphics[width=.99\linewidth,clip=]{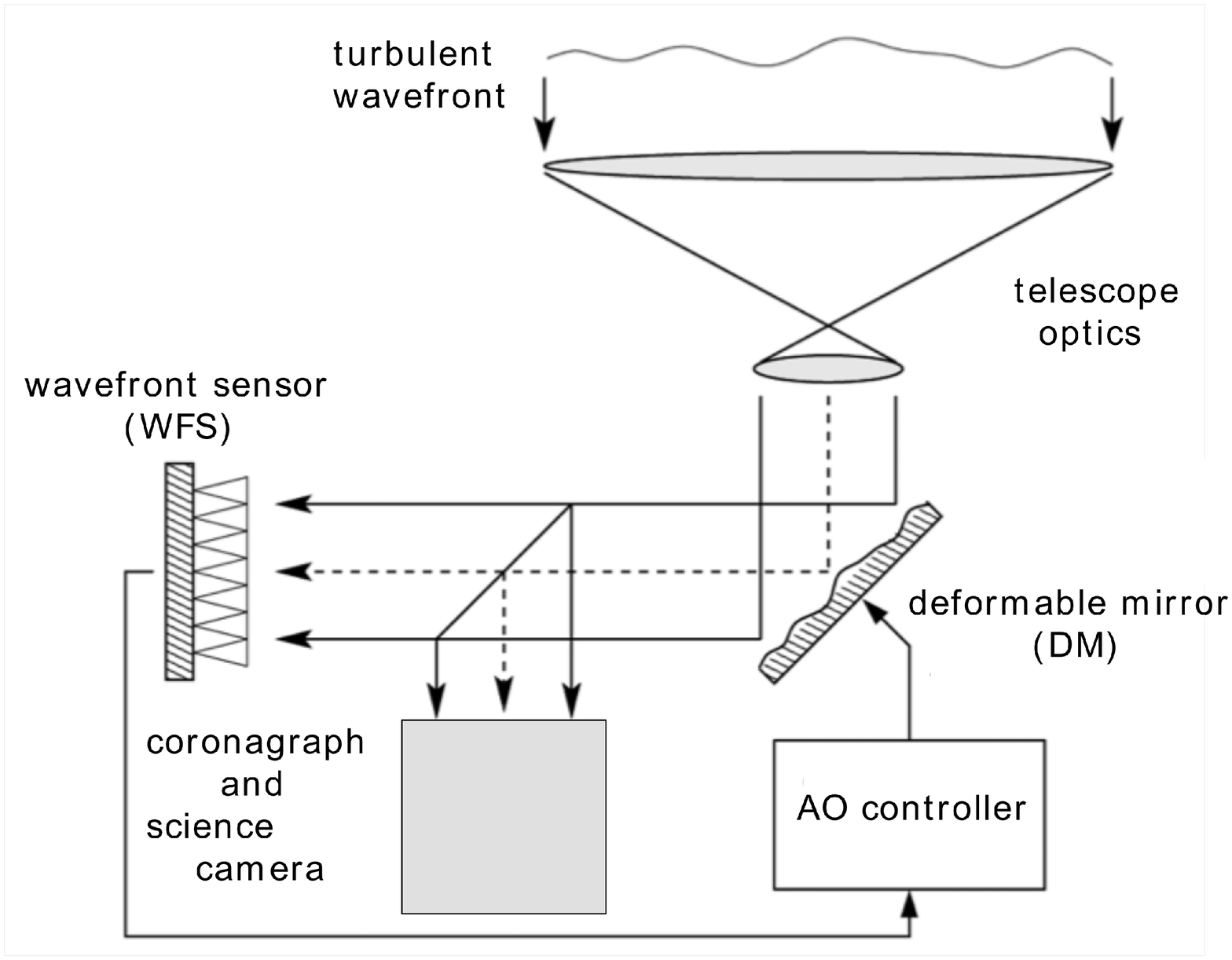}
	\caption{\small Schematic diagram of an astronomical telescope with a closed-loop AO system and a coronagraph.  Modified from [\citenum{Hinnen_H2control}].}
	\label{fig: schematic}
	\vspace{-4mm}
\end{figure}

The assumed observational set-up is depicted schematically in Fig.~\ref{fig: schematic}: there is a ground-based telescope with an adaptive optics (AO) system, and beam splitter or dichroic that sends some of the light to a stellar coronagraph where images are formed on a science camera in a downstream focal plane.
The closed-loop AO system consists of a deformable mirror (DM) and a wavefront sensor (WFS) that measures the phase (and possibly amplitude in future designs) of the corrected wavefront, which we refer to as the \emph{AO residual phase} or simply the \emph{AO residual}.

Briefly, ground-based imaging of exoplanets is made difficult by two sources of aberrations.
The first is the AO residual, which evolves at kHz frequencies due to atmospheric turbulence, but averages out to spatially smooth halo that one could reasonably hope to subtract as a background image.
The second is so-called \emph{non-common path aberrations (NCPA)}, which are hardware aberrations that are not corrected by AO system, usually because they exist in the coronagraph optics train downstream of the beam splitter.
The NCPA create speckles in the coronagraphic focal plane, and are often much brighter than the planets one might hope to see.
The NCPA vary wide range of time-scales as the various mechanical stresses on the optics vary, and are generally considered the primary difficulty in direct imaging of exoplanets.
In classical imaging modes, with exposure times that range from several seconds to many minutes, the speckles caused by the NCPA are rather difficult to distinguish from planets.
(Part I has references to relevant literature.)

Building off the pioneering work of Ref.~[\citenum{Gladysz10}], Frazin was first to demonstrate semi-analytically and in simulation that when the science camera exposure time is about 1 millisecond (ms) or less, the stellar speckle and exoplanets exhibit much different temporal behavior.\cite{Frazin13}
The reason for this is that, at the planet's location in the coronagraph focal plane, the planet's intensity will be proportional to the instantaneous Strehl ratio achieved by the AO system, while the starlight at that same pixel will depend on the random phases in the AO residual that happen to be present.
Thus, the planet's intensity will be far less volatile than the speckles in that pixel.
In that same article, Frazin was the first to propose the possibility of utilizing the WFS telemetry in a regression model to estimate a set of coefficients that simultaneously specify the NCPA and the image of circumstellar material ("the exoplanet image").
(Note that Ref.~[\citenum{Jefferies_WFSdecon13}] uses the WFS telemetry to provide an initial estimate of the blurring function in ground-based satellite imaging.)
The simulations of Ref.~[\citenum{Frazin13}] assumed that the WFS measured the AO residual without error, an obvious simplification used to make the regression model tractable.
Indeed, when one assumes that the AO residual phase values are known without error, the regression model is quite straightforward: it has a mild nonlinearity in the NCPA coefficients, which can be treated iteratively with relinearization, and is linear in the coefficients that specify the exoplanet image.
Introducing wavefront measurement error to problem makes it more difficult, as this article will hopefully make clear. 
The sources of wavefront measurement error include:
\begin{itemize}
	\item insensitivity to certain spatial and temporal frequencies and aliasing
	\item nonlinearity of the intensity measured by the WFS camera in the AO residual phase (and amplitude) values
	\item noise in the intensity measured by the WFS camera.
\end{itemize}
The approach advocated in this article allows these effects to be included in the model of the WFS.

Part I presents simulation results of three regression models, which were called the \emph{ideal estimator}, the \emph{na\"ive estimator} and the \emph{bias-corrected estimator}.
These fairly realistic end-to-end simulations of an AO system with a coronagraph show the potential value of the regression models based on millisecond imaging, with highly accurate estimates of the NCPA and the exoplanet image obtained with only a few minutes of sky time. 
The ideal estimator is essentially Frazin's 2013 regression, which assumes that the true wavefront values are known and, therefore, is not physically realizable.
However, it does serve as a useful benchmark in simulation studies.
To calculate the ideal estimator, one uses the true AO residual wavefronts (known from simulation) in the regression equations presented later in this article.
Note that the ideal estimator does account for noise in the coronagraph images.
Unlike the ideal estimator, the na\"ive estimator is realizable.
The na\"ive estimator makes use of WFS and coronagraph optical models but assumes nothing about the statistics of the AO residuals.
It simply treats flawed WFS measurements as if they were the true values and plugs them into the regression equations for the ideal estimate.
The result is a biased estimate, but the Part I simulations showed that the na\"ive regression based on one minute of simulated sky time estimated an NCPA of $\sim 0.5\,$radian RMS with $\sim 90\%$ accuracy on a magnitude 8 star and $\sim 95\%$ accuracy on a magnitude 6 star.
Thus, despite its simplistic nature, na\"ive estimation may be useful in an NCPA control loop with an update frequency of $\sim 1\,$ minute.

The bias-corrected estimator, whose equations are presented for the first time in this article, assumes complete knowledge of the spatial statistics of the random process responsible for generating the AO residual wavefronts.
This statistical information is used to generate Monte Carlo wavefronts.
The bias-corrected estimator also requires models of the WFS and coronagraph hardware.
If its underlying assumptions are met, the bias-corrected estimator is unbiased in the large sample limit and has error statistics that are quite close to those of the ideal estimator.
While the statistical knowledge will never be complete, approximate knowledge may lead to an estimator that is much less biased than the na\"ive estimator.
In fact, the Part I simulations of the bias-corrected estimate provide an illustrative example of incomplete statistical knowledge.
In those simulations, the phase values of the AO residual wavefronts were not consistent with a multivariate normal distribution (this is discussed later), but the Monte Carlo wavefronts were generated by a multivariate normal with same 2\underline{nd} order statistics as the true AO residual wavefronts.
The bias-corrected regression results were based on 4 minutes of simulated sky time of a magnitude 8 star with an NCPA of $\sim 0.05 \,$radian RMS and a $13 \times 8$ grid of points (located at angular distances ranging from 3 to 10 $\lambda/D$ from the star in the center) representing the exoplanet image.
The bias-corrected estimator simultaneously achieved an NCPA estimate with an RMS accuracy of $\sim 0.004 \,$radian and contrast of $\sim 10^{-5}$ on the image grid.
In addition, the estimate of the exoplanet image is completely free of the self-subtraction artifacts that always plague differential imaging.
Indeed, the estimate of the exoplanet image was nearly identical to the image obtained by subtraction of a perfect point-spread function (PSF).
(One can think of perfect PSF subtraction as the equivalent of ideal estimation for differential imaging.)

In statistical language, the wavefront measurement errors introduce errors into the independent variables of the regression model, thus bringing the problem into the realm of "errors-in-variables" models, an area that has maintained a community of statistical researchers for decades.\cite{Stefanski_Nonlinear}
Na\"ive estimation is a term from the errors-in-variable literature, and it corresponds to simply ignoring the measurement errors in the regression equations.
Ideal estimation is also a term from the errors-in-variables literature, and it corresponds to the hypothetical situation in which the errors in the independent variables are zero.

While some of the ideas presented herein, such as correction of biased estimators, will be  familiar to statisticians, others are original to this article and specific to the problem at hand.
The following presentation draws upon a variety of concepts from statistics and scalar physical optics.
The authors have made every attempt to minimize mathematical formalities and the number of equations, but not at the expense of clarity.

\section{Wavefronts}\label{sec: Wavefronts}

The regression method presented here assumes that the AO system is operating in closed-loop mode, so that the wavefront sensor (WFS) is measuring the wavefront after the DM has applied a correction to compensate for the effects of atmospheric turbulence.
These corrected wavefronts are called the AO residual wavefronts.

The formalism here requires that both the science camera and the WFS are operating at a cadence of roughly a millisecond or less and are taking exposures simultaneously, with exposures from each having the same time-tags $\{ t \}$ (which also serves as the time index).
The total observation time allows $T$ millisecond exposures, so the index $t$ is an integer that runs between $0$ and $T-1$.
Since the exposure time is on the order of a millisecond, it only takes roughly $17$ minutes to achieve $T \approx 10^6$ exposures.
For example, the Phase B simulation in Part I corresponds to a 4 minute observation, corresponding to $T = 2.4 \times 10^5$ exposures.

AO systems with natural guide stars run the wavefront sensors at a frame rate of between 0.5 and 3 kHz.
In practical terms, exposures of a duration of about a millisecond or less are required to effectively capture an instantaneous wavefront.
This paper will assume that the exposure lengths are short enough to effectively freeze the wavefront.
Treating the subtle effects of the slight variation of the wavefront within the duration of one exposure is beyond the scope of this paper and will require subsequent analysis.

Physically, the wavefront represents the phase (and, if necessary, the amplitude perturbations caused by scintillation resulting from multi-layer turbulence) of the electric field in the entrance pupil of the WFS. 
The light is taken to be quasi-monochromatic, centered at the wavelength $\lambda$.
The entrance pupil is modeled as consisting of $P$ pixels, with the phase at the $l$\underline{th} pixel represented as $\phi_l  ( \bw_t )$, so that the complex-valued electric field at that pixel is:
\begin{equation}
u_l'(\bw_t) = \exp[ j  \phi_l  ( \bw_t ) ] \, ,
\label{eq: WFS entrance field}
\end{equation}
where $t$ is the time index, and the phase, $\phi_l(\bw_t)$ will be complex valued if treating the amplitude effects is deemed necessary; if not, it is real-valued.
As an example, the simulations in Part I parameterize a circular aperture inscribed in a $50 \times 50$ pixel grid (the primary mirror diameter is $6.5 \,$m), corresponding to $P=1976$ pixels.

The vector $\bw_t$ is a set of parameters that specify the wavefront over all $P$ pupil pixels.
The forms of the functions $\{ \phi_l(\bw_t) \}$ can be anything that is reasonable, such as splines or Zernike polynomials.
In the Part I simulations, the vector $\bw_t$ also has $P$ components, with each component representing the phase value at the corresponding pixel.
If we were to consider amplitude effects as well, the size of the vector $\bw_t$ would need to increase accordingly.

Once the choice of the $\{ \phi_l \}$  functions has been made, the vector $\bw_t$ fully specifies instantaneous wavefront at time $t$.
Sometimes $\bw_t$ will be referred to as “the wavefront" with the understanding that it is only a parametric representation.
The vector containing the (complex) electric field values at all $P$ pixels in the entrance pupil is $\bu'(\bw_t) = (u_{0}'(\bw_t), \, \dots , \, u_{P-1}'(\bw_t))^\rT$. 
The vector of field values, $\bu'(\bw_t)$, here is denoted with a prime to emphasize that it corresponds to the field entering the WFS, as opposed to to $\bu(\bw_t)$, which is the field entering the coronagraph, with the difference arising due to NCPA.
In the Part I simulations, the vectors $\bu(\bw_t)$ and $\bu’(\bw_t)$ also have 1976 components, due to the pixel-wise representation of the phases in $\bw_t$.

\subsection{Monte Carlo Wavefront Generation}\label{sec: MC generation}

The bias-corrected estimator requires the ability to generate Monte Carlo wavefronts with the same spatial statistical properties as the real wavefronts for the purpose of Monte Carlo calculations.
The set of Monte Carlo wavefronts is denoted by $\{ \breve{\bw}_l \}, \, 0 \leq l  \leq L-1$, where there are $L$ samples in the set.
(In statistics, the process of generating a Monte Carlo data set is called "sampling," so a given Monte Carlo wavefront can be called a "sample.")  
To generate samples of multivariate normal with $P \times 1$ mean vector $\bmu$ and $P \times P$ covariance matrix $\bC$, one simply takes a vector of $P$ samples of a standard normal and multiplies it by a matrix square-root of $\bC$ and adds $\bmu$ to the result.

The true wavefronts, $\bw_0,\; \dots , \; \bw_{T-1} $, form a time series with a characteristic correlation time.
For example, in the Part I simulations, the autocorrelation of the time-series corresponding to the phase of a given pupil pixel drops by a factor of 3 in about 0.013~s and by a factor of about 10 in times ranging roughly between 0.025 and 0.1~s.  
For $T=6 \times 10^4$, corresponding to a 1 minute observation, $T$ is many orders of magnitude greater than any relevant correlation time.  
Unlike the true wavefronts, the $L$ simulated wavefronts $\{ \breve{\bw}_l \}$ form an unordered set.
Under conditions of quasi-stationarity, of which stationarity is a limiting case, this ability to perform Monte Carlo sampling allows approximation of the time average of any function (linear or otherwise) $f(\bw_t)$ with the Monte Carlo mean of $f(\breve{\bw}_l)$.
In other words,
\begin{equation}
\left<  f(\breve{\bw}_l ) \right>_\mathrm{mc} \approx  \left<  f(\breve{\bw}_l ) \right>_\rE
= \left<  f(\bw_t )\right>_\tau \,
\label{eq: mean_mc = mean_E = mean_T}
\end{equation}
which says that (given enough realizations in  $\{ \breve{\bw}_l \}$ ) the Monte Carlo mean of $f(\breve{\bw}_l)$ will closely approximate the ensemble mean  $\left<  f(\breve{\bw}_l \right>_\rE $, which is equal to the temporal mean over the time interval of length $T$ of the stochastic process $ \left<  f(w_l \right>_\tau  $.
The Monte Carlo mean based on the $L$ samples, $\{ \breve{\bw}_l \}$, is:
\begin{equation}
\left<  f(\breve{\bw}_l ) \right>_\mathrm{mc} \; \equiv \frac{1}{L} \sum_{l=0}^{L-1}  f(\breve{\bw}_l ) \, .
\label{eq: mean_mc of f}
\end{equation}

\subsection{Wavefront Measurements}

The WFS telemetry makes possible an estimate (or measurement) of $\bw_t$, denoted as $\hat{\bw}_t$, where the "hat" $\hat{\bw}_t$ indicates that this vector is an estimate.
In the Part I simulations, the vector $\hat{\bw}_t$ representing the measured wavefront also has 1976 components.
It will be assumed that WFS can be simulated accurately, which is an added prerequisite to on-sky implementation of this algorithm.
The wavefront estimate, also called the "wavefront measurement," is related to the true wavefront via:
\begin{equation}
\hat{\bw}_t = \mathcal{W}(\bw_t) + \bn_t \, ,
\label{eq: WFS measurement}
\end{equation}
where $\mathcal{W}$ is a measurement operator that is nonlinear, and the noise vector is denoted as $\bn_t$.
$\mathcal{W}$ is necessarily nonlinear due to the fact the intensity measured by the detector in the WFS is nonlinear in the phase values.
Thus, this nonlinearity is inevitable even if $\hat{\bw}$ is based on an estimator that is linear in the WFS intensity values.
Numerically, $\mathcal{W}$ is a realization of the measurement operator, meaning that there is a computer program that takes an input wavefront vector $\bw_t$, and simulates the WFS optical train as a well as the needed signal processing to produce an estimate the wavefront, $\hat{\bw}_t$.
A previous article by Frazin provides linear and nonlinear estimation algorithms for the case of a pyramid wavefront sensor,\cite{Frazin_JOSAA2018} but the formalism here is completely agnostic with respect to the wavefront sensing hardware.

Given the numerical realization of the wavefront measurement operator $\mathcal{W}$ and a Monte Carlo sample $\breve{\bw}_l$, it is straightforward to sample from the distribution of the wavefront estimates:
\begin{equation}
\hat{\breve{\bw}}_l = \mathcal{W}(\breve{\bw}_l) + \bn_l \, ,
\label{eq: simulated WFS measurement}
\end{equation}
where $\bn_l$ is a sample from the same stochastic processing governing $\bn_t$ in \eqref{eq: WFS measurement}.

\section{Coronagraph and NCPA Model}
\subsection{Stellar Intensity}

The formalism given here is agnostic with respect to the hardware in the coronagraph optical train.
Indeed, there is nothing in the mathematical formalism itself that requires coronagraphic hardware.
(For that matter, the demonstrated on-sky starlight suppression of coronagraphy from ground-based platforms is less than impressive, perhaps a factor of 2 at distances beyond about 2 $\lambda/D$ from the star [see, e.g., Fig.~9 of Ref.~\citenum{Guyon_ExAOreview18}].)
The field in the coronagraph entrance pupil is the same as that in the WFS entrance pupil, as given in \eqref{eq: WFS entrance field}, except that is it modified by the NCPA, and it is given by:
\begin{equation}
u_l(\bw_t, \ba) = u_l'(\bw_t) \exp[ j  \theta_l  ( \ba) ] = \exp j[ \phi_l  ( \bw_t ) +  \theta_l  ( \ba) ] \, ,
\label{eq: coronagraph entrance field}
\end{equation}
where $u_l'(\bw_t)$ was given in \eqref{eq: WFS entrance field}, and $\ba$ is a vector of length $N_\ra$ containing real-valued parameters that specify the NCPA.
Sometimes $\ba$ will be referred to as "the NCPA", which is shorthand for  "the NCPA parameters."
The function $ \theta_l  ( \ba) $ expresses the phase (and, if needed, the amplitude as well) of the NCPA at the $l$\underline{th} pixel in the pupil as a function of the vector $\ba$.
The simplest way to do this to have $a_l$, which is the $l$\underline{th} component of $\ba$, be the phase of the NCPA at the $l$\underline{th} pupil pixel so that $\theta_l(\ba) = a_l$, but other representations, such as modal expansions, may be useful and require fewer components in the $\ba$ vector.
In the Part I simulations, the NCPA is represented by 33 Zernike modes (up to 6\underline{th} radial order without the piston, tip or tilt terms), so the vector $\ba$ has 33 components.
The vector containing the field values at all $P$ pupil pixels is $\bu(\bw_t, \ba) = \big(u_0(\bw_t, \ba), \, \dots , \, u_{P-1}(\bw_t,\ba) \big)^\rT$.

The estimation methods presented here are not limited to pupil-plane manifestations of the NCPA, although that is all that is treated here.
Adding additional quantities to be estimated from the regression, such as coronagraph alignment errors or NCPA that are not accurately described by pupil plane manifestations is also possible by augmenting the regression model appropriately.

It is critical to keep in mind that, $\theta_l(\ba)$, which is the phase (and possibly amplitude) perturbation caused by the NCPA is intended to model hardware aberrations that fluctuate on the timescales at which the various mechanical stresses vary.
To keep the regression models from becoming more complicated than they already are, we have assumed that NCPA do not change over the observation time, $T$ (which was only 1 or 4 minutes in the Part I simulations).
In principle, the regression models presented here can be placed within a Kalman filtering approach to allow time-dependence of the NCPA.
Indeed, including rapid vibrations is not out of the question, especially if the power of the vibrations is somehow confined in time-frequency space.
(Vibrations are often sensed with accelerometers.  
For example, the Subaru telescope exhibits intermittent periods having vibrations with frequencies up to about 30 Hz.\cite{Lozi_vibrations_JATIS18}
In Ref.~[\citenum{Frazin14}], Frazin demonstrated ideal estimation of a vibration amplitude with a known frequency.)

Since the coronagraph is a linear optical system (i.e., it does not contain active optical elements such as gain media as lasers do), the field at the $n$\underline{th} detector pixel in the science camera is a linear combination of the field values at the pupil pixels.
 Then, the vector containing the detector field values at time $t$ is given by the matrix-vector multiplication:
 \begin{equation}
 \bv(\bw_t, \ba) = \bD \, \bu(\bw_t, \ba) \,
 \label{eq: coronagraph field prop}
 \end{equation}
where $\bv(\bw_t, \ba)$ has $M$ components, one for each of the $M$ science camera pixels, and $\bD$ is a complex-valued matrix that performs the optical propagation.
The  $\bD$ matrix of size $M \times P$ is the result of pre-computed optical propagations.
In the Part I simulations, the science camera had $M = 67 \times 67 = 4489$ pixels and there were $P=1976$ pupil pixels, so the $\bD$ matrix was $4489 \times 1976$.
Pre-computing the matrix $\bD$ allows computationally expensive modeling to be employed at no cost to the execution time of this regression algorithm.
An analogous procedure for pre-computing the propagation matrix for a pyramid WFS is given in \cite{Frazin_JOSAA2018}.
The stellar component of the intensity in the science camera at time $t$ is then
\begin{equation}
\bi_\star (\bw_t, \ba) = \bv(\bw_t, \ba) \circ \bv^*(\bw_t, \ba) \, ,
\label{eq: star intensity}
\end{equation} 
where the $\circ$ notation denotes element-wise multiplication, otherwise known has a Hadamard product.

\subsection{Planetary Intensity}

In this section an expression for the instantaneous planetary intensity impinging on the detector at time $t$ is provided.
This intensity will vary in time because the planetary light rays experience the same wavefront perturbations, $\bw_t$, that the starlight does.
Implicit in this assumption is that the effects of anisoplanatism are negligible, which will certainly be the case since the separations between the sought-after targets and the star are much less than an arcsecond.\cite{Fried_anisoplanatism82}
In order to simplify the discussion a bit, it will also be assumed that the effects of NCPA are negligible as far as the planetary light is concerned.
This is probably a valid approximation, but including the NCPA in the present formulation is quite straightforward, requiring linearization in the NCPA coefficients $\ba$, which needs to be done anyway for the starlight, as shown in Subsection~\ref{sec: linearization}.
Under the standard assumption that astronomical sources are spatially incoherent, which is basis of the famous Van Cittert - Zernike theorem,\cite{StatisticalOptics} the planetary intensity impinging on the science camera is linear in the fluxes coming from the elementary patches of solid-angle on the sky.
This linearity makes this regression modeling  task a bit easier, as the resulting regression model is linear in the parameter vector $\bp$ that represents the planetary image.
Sometimes the vector $\bp$ will be referred to as the \emph{planetary image}, even though it is a set of parameters that specifies the image, not the image itself.

Let the sky-angle relative to the telescope pointing be represented by the two-component vector $\balpha = (\alpha_x, \, \alpha_y)$, where $\alpha_x$ and $\alpha_y$ correspond to local Cartesian coordinates on the sky, and these angles have units of radians.
This local Cartesian coordinate system is perfectly acceptable due to the sub-arcsecond angles involved.
Physically, the value of $| \balpha |$ will range from about  $0.5$ to perhaps $10$ times $\lambda/D$.
The image we wish to estimate is represented by the function $S(\balpha)$, which has units of energy flux per solid angle.
Note that $S(\balpha)$ excludes the star and only represents the surrounding planetary system. 
Let $\{ \balpha_n \}$, correspond to a numerical grid of angles, with $N_p$ points on the sky that we will use to represent the image $S(\balpha)$, so that
\begin{equation}
S(\balpha) = \gamma \sum_{n=0}^{N_p-1} p_n \delta(\balpha - \balpha_n)
\label{eq: planetary image expansion}
\end{equation}
where $\gamma$ contains the necessary scaling factors, $N_p$ is the number of points in the grid, the coefficients $\{ p_n \}$ are unitless, and $\delta $ is the Dirac delta function.

The expansion \eqref{eq: planetary image expansion} should be adequate if the grid spacing of the angles $\{ \balpha_n \}$ is roughly $\lambda/D$ or less.
The Part I simulations used a 1~$\lambda/D$ grid spacing.
Preliminary calculations indicate that spacing as small as perhaps $0.7 \, \lambda/D$ (corresponding to mild superresolution) may be practical without incurring a large penalty in the estimate error. 
Next, it is useful to define the vector $\bp = (p_0, \, \dots \, \, p_{N_p-1})^\rT$, which has $N_p$ components, one for each sky-angle on the grid. 
In the Part I simulations, the planetary image was specified on a $13 \times 8$ $(\alpha_x, \alpha_y)$ grid, so $\bp$ had $N_p = 104$ components.
Other representations, such as spline basis functions, may be more efficient in terms covering the same portion of sky with fewer parameters, but that is a detail that need not be covered here.
Mathematically, the essential property of the expansion in  \eqref{eq: planetary image expansion} is that it is linear in the coefficients $\{ p_n \}$.

In the telescope beam above the atmosphere, at the transverse location $\br$, the electric field from the portion of the planetary image at the angle $\balpha_n$ is given by:
$ u_{\rp n}(\br) = \sqrt{p_n} \exp j [ 2 \pi \balpha_n \cdot \br / \lambda   ]$ where $\lambda$ is the wavelength and $\balpha_n \cdot \br$ is a scalar ("dot") product.
After passing through the atmosphere and the telescope's pre-coronagraph optics (including the AO system), the field arising from the sky-angle $\balpha_n$ at time $t$ and location $\br_l$\ in the coronagraph entrance pupil is given by:
\begin{equation}
u_{\rp t, n,  l} = \sqrt{p_n}  \exp j [ \phi_l(\bw_t) +  2 \pi \balpha_n \cdot \br_l / \lambda  ] \, ,
\label{eq: u_p}
\end{equation}
where the AO residual, $\phi_l(\bw_t)$, has been included and the angle $\balpha_n$ has been tacitly rescaled to account for the magnification of the telescope's beam reducing optics.
The reader is reminded that the effect of the NCPA has been neglected, as per the above remarks.
It is now convenient to collect the fields defined by \eqref{eq: u_p} at all $P$ pixels in the pupil into a vector (still coming from the one sky-angle $\balpha_n$), defining
the vector $\bu_{\rp t, n} = ( u_{\rp t, n, 0} \; , \:  \, \dots \: , \;  u_{\rp t,  n, P-1} )^\rT /  \sqrt{p_n}$,  where the amplitude,   $\sqrt{p_n} $, has been normalized out for reasons that will become obvious shortly.
Similarly to \eqref{eq: coronagraph field prop}, the $\bD$ matrix can be used to propagate this planetary field to the detector plane, defining the vector $\bv_{\rp t , n}$: 
\begin{equation}
\bv_{\rp t , n} =  \sqrt{p_n} \, \bD \, \bu_{\rp t, n}
\label{eq: planet field prop}
\end{equation}
where $\bv_{\rp ,t, n}$ is a vector of length $M$ (one component for each detector pixel) containing the field in the detector plane arising from the part of the planetary image at the sky-angle $\balpha_n$.
Next, the intensity impinging on the $M$ detector pixels due to the part of the image at $\balpha_n$ at time $t$ is given by the vector $\bi_{\rp,n}$:

\begin{equation}
\bi_{\rp t, n} = \bv_{\rp t , n} \circ \bv^*_{\rp t , n}   = p_n (\bD \bu_{\rp t, n }) \circ (\bD^* \bu^*_{\rp t, n }) \, .
\label{eq: planetary intensity from n}
\end{equation}
The vector $\bi_{\rp t, n}$ is the instantaneous image of a planet at the location $\balpha_n$ on the science camera.
Since the planetary intensity  can be considered to be the sum of incoherently radiating sources from the directions $\{ \balpha_k \}$, to get the instantaneous planetary image at time $t$, we only need to sum $\bi_{\rp t, n }$ over the $n$ sky-angles:
\begin{align}
\bi_\rp(\bw_t, \bp) = \sum_{n=0}^{N_p-1} \bi_{\rp t, n} & =  \sum_{n=0}^{N_p-1} 
p_n (\bD \bu_{\rp t, n }) \circ (\bD^* \bu^*_{\rp t, n })  \nonumber \\
& \equiv \bA_\rp(\bw_t) \bp \, , 
\label{eq: planetary intensity}
\end{align}
where $\bp$ is the vector of planetary image coefficients and  $ \bA_\rp(\bw_t) \bp $ is a matrix-vector multiplication.
\eqref{eq: planetary intensity} defines the real-valued, $M \times N_p$  planetary system matrix as $ \bA_\rp(\bw_t) \equiv (\bD \bu_{\rp t, n }) \circ  (\bD^* \bu^*_{\rp t, n }) $.
The dependence of $ \bA_\rp(\bw_t)$ on the $M$ detector pixels and the $N_p$ sky-angles can be seen in \eqref{eq: u_p}.
The planetary system matrix, $ \bA_\rp(\bw_t) $ is not a constant matrix, rather is a matrix of nonlinear functions of the instantaneous wavefront parameters $\bw_t$.
To put it another way, $\bA_\mathrm{p}(\bw_t)$ provides for the modulation of the planetary image by the AO residual.

\subsection{Linearization}\label{sec: linearization}

The regression method presented below relies on linearity in the parameters to be estimated, which include the NCPA coefficient vector $\ba$.
The simulations in Part I show that 3 cycles of linearized estimation followed by re-estimation have no problem estimating NCPA with RMS phase of $\sim 0.5\,$radian.
These and other unpublished simulations indicate there is no reason to believe that that nonlinearity in the NCPA coefficients limits the accuracy of the method for any reasonable amplitude of the NCPA.
Thus, we will proceed on premise that linearizing the problem in the coefficients $\ba$ is an acceptable procedure.
Note that this procedure of successive relinearizations is similar to gradient descent to handle nonlinearity in the regression problem.
Importantly, the estimators presented here never linearize in the wavefront $\bw_t$, the measured wavefront $\hat{\bw}_t$, nor even the wavefront measurement error $\delta \bw_t \equiv \bw_t -\hat{\bw}_t$.

To get the total intensity impinging on the detector, we sum the stellar intensity from \eqref{eq: star intensity} with the planetary intensity from \eqref{eq: planetary intensity}:
\begin{align}
 \bi(\bw_t, \ba, \bp) &\equiv \bi_\star (\bw_t, \ba) + \bi_\rp(\bw_t, \bp) \nonumber  \\ 
  & = \bi_\star (\bw_t, \ba) + \bA_\rp(\bw_t) \bp
\label{eq: total intensity}
\end{align}
A Taylor expansion of  $\bi_\star (\bw_t, \ba) $ in \eqref{eq: total intensity} in the vectors $\ba$ and $\bp$ about the point $(\ba_0, \bp_0)$ can be written as:
\begin{equation}
\bi (\bw_t, \ba, \bp) \approx \bc(\bw_t) +  \bA_\ra(\bw_t)  (\ba - \ba_0) 
+  \bA_\rp(\bw_t) (\bp - \bp_0) \, ,
\label{eq: total Taylor}
\end{equation}
where $ \bA_\ra(\bw) $ is the real-valued $M \times N_\ra$ \emph{NCPA system matrix}.
It is a matrix of functions of $\bw_t$ defined as the Jacobian:
\begin{equation}
 \bA_\ra(\bw_t)  \equiv \frac{\partial \, \bi_\star(\bw_t, \ba) }{\partial \ba} \bigg{|}_{\ba_0} \, ,
\label{eq: NCPA system matrix}
\end{equation} 
which can be calculated from \eqref{eq: star intensity}.
The zero point vector is defined as:
\begin{equation}
 \bc(\bw_t) \equiv \bi (\bw_t, \ba_0, \bp_0) \, .
 \label{def: c(w_t)}
\end{equation}

In order to avoid carrying unnecessary notation (as most readers will surely welcome this), without loss of generality, the vectors $\ba_0$ and $\bp_0$ will be taken to be 0.
In addition, since the iterative procedure re-linearization and estimation has proved so successful in simulation, the approximation in \eqref{eq: total Taylor} will be taken to be an equality, which will also help to make the following discussion less complicated.
Then, for the purposes of the rest of this paper, we have the following expression for the instantaneous total intensity impinging on the science camera:
\begin{equation}
\bi (\bw_t, \ba, \bp)  = \bc(\bw_t) +  \bA_\ra(\bw_t)  \ba + \bA_\rp(\bw_t) \bp  \, ,
\label{eq: total intensity reduced}
\end{equation}
which models all of the speckles, planetary and otherwise, seen on the science camera at time $t$.
Note that $\bi(\bw_t, \ba, \bp)$ does not refer to the quantities measured by the science camera, since it does not include noise.
Rather, even within the paradigm of this model (i.e., pretending the above model is exactly correct), it is a quantity that can never be known exactly since $\bw_t$, $\ba$ and $\bp$ can only be estimated.
A similar equation was given by Ref.~[\citenum{Sauvage_model10}] early in the phase diversity literature, although they were concerned with long-exposure imaging.

\subsection{Setting Up the Regression}

The discussion of the regression method presented below is lengthy, and a number of new variables are assigned that will economize the notation as well as provide clarity.
The first such step is defining the vector $\bx$ with $N = N_\ra + N_\rp$ components, which concatenates $\ba$ and $\bp$, and the  $M \times N$ matrix $\bA(\bw_t)$:
\begin{align}
\bx & \equiv \left(\begin{array}{c} \ba \\ \bp \end{array}\right) \, \: \: \mathrm{and} \:
\label{def: x} \\
\bA(\bw_t)  & \equiv \bigg( \bA_\ra(\bw_t) \: \: \: \: \bA_\rp (\bw_t) \bigg) \, .
\label{def: x A}
\end{align}
With these definitions, \eqref{eq: total intensity reduced} can be written more compactly as:
\begin{equation}
\bi(\bw_t, \bx) =   \bc(\bw_t) +  \bA(\bw_t) \bx \, .
\label{eq: total intensity}
\end{equation}
The vector $\by_t$, which has length $M$, represents the intensities measured by the science camera pixels at time $t$.
It differs from $\bi(\bw_t, \bx)$ due to shot-noise, readout noise, thermal background and other effects that need to be included.
$\by_t$ is given by the equation:
\begin{align}
\by_t & = \bi(\bw_t, \bx) + \bnu_t  \nonumber \\
& =   \bc(\bw_t) +  \bA(\bw_t) \bx  + \bnu_t  \, ,
\label{eq: y_t}
\end{align}
where the vector $\bnu_t$ (also of length $M$) represents noise in the measurements, which are taken to be samples of zero-mean stochastic process with $M \times M$ covariance matrix $\bC_t$.
Note that while shot-noise is significantly non-Gaussian in the low-count regime, it always has a well-defined covariance matrix.
As concerns the regression, $\by_t$ is a fixed vector of measured values that have been provided by the science camera. 
The goal of the regression is, of course, to estimate $\bx$.

If the science camera and WFS are running at 1 kHz, the number of exposures, $T$, reaches $10^6$ in under 17 minutes, and an astronomical observation of a single exoplanet target may be hours in duration.
Thus, it is necessary to develop a regression framework that makes statistical inference based on large numbers of exposures practical.
The vectors $\bw$, $\by$ and $\bnu$ are defined as concatenations of their namesakes:
\begin{equation}
\bw \equiv 
\left(\begin{array}{l} \bw_0 \\  \vdots \\ \bw_{T-1} \end{array}\right) \, ,
\:
\by \equiv 
\left(\begin{array}{l} \by_0 \\  \vdots \\ \by_{T-1} \end{array}\right) \, ,
\:
\bnu \equiv 
\left(\begin{array}{l} \bnu_0 \\  \vdots \\ \bnu_{T-1} \end{array}\right) \, ,
\label{def: y nu}
\end{equation}
It will be assumed the covariance matrix of $\bnu$, $\bC$, consists of diagonal blocks $\bC_t$.
The vectors $\by$ and $\bnu$ have lengths $MT$.
The lengths $\bw_t$, and $\bw$ are not specified here because the manner in which the parameter vector $\bw_t$ represents the phase has been left open for the purposes of this discussion.
For reference, in the Part I simulations, the length of $\bw_t$ was 1976 (each value being the phase of a pupil pixel), so that the length of $\bw$ was $1976T$ (values of $T$ corresponded to 1 and 4 minutes were used in the simulations).
Similarly, the following concatenations are defined:
\begin{equation}
\bd(\bw) \equiv 
\left(\begin{array}{l} \bc( \bw_0) \\  \vdots \\ \bc(\bw_{T-1}) \end{array}\right) \, ,
\:
\bZ(\bw) \equiv 
\left(\begin{array}{l} \bA( \bw_0) \\  \vdots \\ \bA(\bw_{T-1}) \end{array}\right) \, ,
\label{def: c Z}
\end{equation} 
where the $MT \times N$ matrix of functions $\bZ(\bw)$ is called the \emph{grand system matrix}.
The measured wavefronts at time $t$ are represented by the vector $\hat{\bw}_t$.  Similarly to $\bw$, the vector $\hat{\bw}$ is defined as the concatenation of the $T$ measured wavefronts:
\begin{equation}
\hat{\bw} \equiv 
\left(\begin{array}{l} \hat{\bw}_0 \\  \vdots \\ \hat{\bw}_{T-1} \end{array}\right) \, .
\label{def: hat w}
\end{equation}
With the definitions in Eqs. (\ref{def: y nu}) and (\ref{def: c Z}), concatenation of \eqref{eq: y_t} over the $T$ exposures is:
\begin{equation}
\by = \bZ(\bw) \bx + \bd(\bw)  + \bnu \, .
\label{eq: y}
\end{equation}
Eq.~(\ref{eq: y}) specifies the measured intensities $\by$ in terms of the true wavefronts $\bw$ and the vector containing the parameters that represent the NCPA and exoplanet image $\bx$. 
In the 4 minute simulation shown in Part I, $T$ was $2.4 \times 10^5$, and the detector had 4489 pixels, so the length of $\by$ was about $1.08 \times 10^{9}$, however, the algorithms work sequentially, making the memory requirements modest.
Those simulations also had 33 NCPA coefficients and 104 image pixels, so $\bx$ had $N = 137$ components.
Given the measured values $\by$, the problem that confronts us can be compactly represented as finding an estimators for $\bx$ in each of the three following situations:
\begin{itemize}
\item{Ideal estimation: the true values $\bw$ are known exactly.}
\item{Na\"ive estimation: the measured values $\hat{\bw}$ are known, but no statistical knowledge of the properties of $\bw$ is provided.}
\item{bias-corrected estimation:  the measured values $\hat{\bw}$ are known, as are the statistical the properties of $\bw$.}
\end{itemize}

\section{Regression}\label{sec: regression}

\subsection{Ideal Estimation}\label{sec: ideal}

The starting point for this analysis is so-called \emph{ideal estimation,} which comes from the errors-in-variables literature.
The ideal estimate is not physically realizable since it requires perfect knowledge of the independent variables (in this case, $\bw$), as is the usual assumption in classical regression.
Despite the fact that ideal estimation is not physically realizable, the ideal estimate makes an appearance in the derivation of the bias-corrected estimate (which is realizable).
Regularized least-squares is the standard way to solve the linear regression problem and it is optimal in Gaussian noise if the measurement weights are chosen to be $\bC^{-1}$.
We will proceed under the assumption that if we were able to realize the ideal estimator it would provide an acceptable estimate.
In terms of the vector we wish to estimate, $\bx$ (which, we remind the reader is a set of coefficients that jointly specify the exoplanet image and the NCPA), the regularized least-squares cost function corresponding to \eqref{eq: y} is:
\begin{align}
\Phi(\bx) = \frac{1}{2}& \big[\bZ(\bw) \bx + \bd(\bw) - \by]^\rT \bS  \big[\bZ(\bw) \bx + \bd(\bw) - \by] 
\nonumber \\ \: \: \: 
& + \: \frac{\beta}{2} (\bx - \bx_0)^\rT \bXi  (\bx - \bx_0) \,
\label{eq: ideal cost}
\end{align}
where $\bx_0$ allows non-centered regularization, $\beta > 0$ is a regularization parameter, $\bXi$\ is a $N \times N$ symmetric (positive semi-definite) reglarization matrix,  $\bS$ is a $MT \times MT$  matrix of measurement weights with the $M \times M$ matrices $\{ \bS_t \}$ on the diagonal blocks, i.e.:
\begin{equation}
\bS = \left( \begin{array}{ccc}
\ddots & \boldsymbol{0} & \boldsymbol{0} \\
\boldsymbol{0} & \bS_t & \boldsymbol{0} \\
\boldsymbol{0} & \boldsymbol{0} & \ddots
\end{array} \right)  \: .
\label{eq: S_t blocks}
\end{equation}
Note that $\bx_0$ in \eqref{eq: ideal cost} is unrelated to the linearization point $\ba_0$ in \eqref{def: c(w_t)}.
Common choices for the regularization matrix include the identity matrix and finite-difference formulations that penalize gradients in the solution.
Standard methods for choosing the regularization parameter include simulation and cross-validation (Frazin does not advocate the "L-curve" method, but that is another matter).
The quadratic form of the regularization term corresponds to a Gaussian prior density (centered on $\bx_0$) in Bayesian inference.
Least-squares, regularization, Bayesian inference, and related matters are covered in many standard references and texts, [e.g., \citenum{Demoment89, Moon&Stirling}], and we will not dwell on them further.

Of course, without knowing $\bw$, there is no obvious way to minimize $\Phi(\bx)$ in \eqref{eq: ideal cost}.
Nevertheless, in this section will proceed as if we did know $\bw$.
The value of $\bx$ that minimizes $\Phi(\bx)$ is called the ideal estimate, $\hat{\bx}_\ri$, and a few lines of algebra show that:
\begin{align}
\hat{\bx}_\ri \equiv 
 \big[ \bZ^\rT(\bw) \bS \bZ(\bw)  & + \beta \bXi \big]^{-1}  \nonumber \\
 & \times \: \big\{ \bZ^\rT(\bw) \bS  [\by - \bd(\bw) ]   + \bXi \bx_0 \big\}  \, .
\label{def: x ideal -1}
\end{align}
In order to economize the discussion, henceforth it will be assumed that $\bx_0 = 0$.
Including $\bx_0 \neq 0$ in the various estimators presented below is straightforward.

For later use, the following definitions are useful:
\begin{align}
\bQ(\bw) & \equiv \bZ^\rT(\bw) \bS \bZ(\bw) \: \: \: \mathrm{and}
\label{def: Q} \\
\bP(\bw) & \equiv  \big[ \bQ(\bw)    + \beta \bXi \big]^{-1} \, .
\label{def: P}
\end{align}
Both $\bQ(\bw)$\ and $\bP(\bw)$\ are $N \times N$ matrices, and $\bP(\bw)$\ is the regularized inverse of $\bQ(\bw)$.

Using Eqs.~(\ref{def: Q}) and (\ref{def: P}), the ideal estimate $\hat{\bx}_\ri$ in \eqref{def: x ideal -1} is written as:
\begin{equation}
\hat{\bx}_\ri = \bP(\bw)   \bZ^\rT(\bw) \bS  [\by - \bd(\bw) ]    \, .
\label{def: x ideal}
\end{equation}
Note that the ideal estimate inherits the noise in measurements $\by$.
It is "ideal" in sense that it does not suffer from the lack of knowledge of the independent variables $\bw$.

Averaging over the statistics of $\bnu$, it is clear from \eqref{eq: y} that $\left< \by \right>_{\bnu} = \bZ(\bw) \bx + \bd(\bw)$, so that from \eqref{def: x ideal}, the mean of $\hat{\bx}_\ri$ is:
\begin{align}
\left< \hat{\bx}_\ri \right>_{\bnu} & =  \bP(\bw) \bQ(\bw) \bx 
\label{eq: mean ideal regularized} \\
& \overset{\beta=0}{=} \bx   \, ,
\label{eq: mean ideal}
\end{align}
indicating the usual result that the ideal estimator is unbiased when no regularization is applied (i.e., $\beta = 0$).
Similarly, the covariance of the ideal estimator is:
\begin{align}
< \hat{\bx}_\ri \hat{\bx}_\ri^\rT >_{\bnu} & - < \hat{\bx}_\ri >_{\bnu} < \hat{\bx}_\ri^\rT >_{\bnu} \nonumber \\
& = \; \bP(\bw)  \bZ^\rT(\bw) \bS \bC \bS \bZ(\bw)  \bP(\bw)
\label{eq: eq: eq: ideal cov full} \\
& = \; \bP(\bw) \bH(\bw)  \bP(\bw)
\label{eq: ideal cov H} \\
& \overset{\beta=0, \: \bS=\bC^{-1}}{=}  \; \bP(\bw)  \, \,
\label{eq: ideal cov no reg} 
\end{align}
where \eqref{eq: ideal cov no reg} shows the considerable simplification attained when the measurement weights $\bS$ are chosen to be the inverse of the noise covariance (i.e., $\bS = \bC^{-1}$) and no regularization is applied.
In Eq.~(\ref{eq: ideal cov H}), the $N \times N$ matrix $\bH(\bw)$ is defined (for later use) as:
\begin{equation}
\bH(\bw) \equiv \bZ^\rT(\bw) \bS \bC \bS \bZ(\bw) \, .
\label{eq: H def}
\end{equation}
Henceforth, it will be assumed that the weight matrices $\{ \bS_t \}$, the regularization parameter and matrix $\beta$ and $\bXi$ are chosen so that ideal estimator is acceptably biased.
Our shorthand term for "acceptably biased" will be "unbiased," so that the ideal estimator is "unbiased."  

We close our discussion of the ideal estimator with the following helpful observation, which follows from Eqs. (\ref{def: c Z}) and (\ref{def: Q}).
\begin{align}
 \bQ(\bw)  
 & = \sum_{t=0}^{T-1} \bA^\rT(\bw_t) \bS_t \bA(\bw_t) \nonumber \\
 & = T  \left< \bA^\rT(\bw_t) \bS_t \bA(\bw_t) \right>_\tau \, \nonumber \\
 & \equiv \bQ_\tau
 \label{eq: Q is a time average}
\end{align}
which states the matrix $\bQ(\bw)$ is a time average, over the interval defined by the $T$ exposures, of a nonlinear function of the wavefronts $\{ \bw_t \}$.
This finite-time average (times $T$) defines the matrix $\bQ_\tau$.
When $T$ is much greater than the correlation time-scale of $\bw_t$ (see Sec.~\ref{sec: Wavefronts}.\ref{sec: MC generation} for an example of the relevant time-scales), $\bQ_\tau$ essentially depends only on the statistical properties of the wavefronts, not their individual values.
From Eqs. (\ref{def: P}) and (\ref{eq: Q is a time average}), we define:
\begin{equation}
\bP_\tau  \equiv  \left[ \beta \bXi +  \bQ_\tau \right]^{-1}  \, .
\label{eq: P is a time average}
\end{equation}
Using Eqs. (\ref{eq: mean ideal regularized}), (\ref{eq: Q is a time average}) and (\ref{eq: P is a time average}), the mean of the ideal estimator can be expressed in terms of the matrices $\bQ_\tau$ and $\bP_\tau$:
\begin{equation}
\left< \bx_\ri \right>_{\bnu} =  \bP_\tau \bQ_\tau \bx \: 
 \overset{\beta=0}{=} \bx   \, .
\label{eq: approx mean ideal regularized}
\end{equation}

\subsection{Na\"ive Estimation}

\emph{Na\"ive estimation} is another term from the errors-in-variables literature.
Na\"ive estimation simply ignores the measurement errors in the independent variables.
To calculate the na\"ive estimate, one replaces true quantities, $\bw$, with the measured ones, $\hat{\bw}$, in the formula for the ideal estimate, which makes it realizable.
The na\"ive estimate, $\hat{\bx}_\rn$, is given by:
\begin{equation}
\hat{\bx}_\rn \equiv \bP(\hat{\bw})   \bZ^\rT(\hat{\bw}) \bS  [\by - \bd(\hat{\bw}) ]    \, ,
\label{def: x naive}
\end{equation}
which is the same as \eqref{def: x ideal}, except that the estimated wavefronts ($\hat{\bw}$) have replaced the true wavefronts ($\bw$).
The simulations given in Part I used na\"ive estimation to good effect in estimating an NCPA with $\sim 0.5\,$radian RMS based on only 1 minute of simulated sky time.
However, when more precision was required, as when jointly estimating the exoplanet image with an NCPA of $\sim 0.05\,$radian with 4 minutes of simulated sky time, the na\"ive estimate was too biased to be useful.
To begin to understand the sources of bias in the na\"ive estimator, one can perform a painstaking Taylor expansion of \eqref{def: x naive} in the wavefront measurement error, $\hat{\bw} - \bw $, and see the many nonlinearities that arise.

\subsection{Biased Estimation}

The particular biased estimator described in this section appears to be a new idea.
We will call this estimator the \emph{biased estimator} (even though the na\"ive estimator is biased, too) in this article.
The only purpose of the biased estimator is as a building block of the bias-corrected estimator.  
This biased estimate, $\hat{\bx}_\mathrm{b}$, is defined as:
\begin{equation}
\hat{\bx}_\mathrm{b} \equiv \bP_\tau   \bZ^\rT(\hat{\bw}) \bS  [\by - \bd(\hat{\bw}) ]    \, .
\label{def: x biased}
\end{equation}
The biased estimate is similar to the na\"ive estimate, except $\bP_\tau$ replaces $\bP(\hat{\bw})$, which removes some of the dependence on the measured wavefronts $\hat{\bw}$.
The biased estimator is not directly useful (as far as the authors can tell).
However, having a Monte Carlo proxy for $\bP_\tau$ (see below) will make our path navigable.

Now, we calculate the bias in the biased estimate so that we can later remove it with Monte Carlo methods.
The wavefront measurement error is defined as:
\begin{equation}
\delta \bw \equiv \hat{\bw} - \bw \, ,
\label{def: wavefront error}
\end{equation}
where the sign is chosen for mathematical convenience.
Following the same sign convention, the following errors are defined:
\begin{align}
\delta \bZ(\hat{\bw}, \bw) & \equiv \bZ(\hat{\bw}) - \bZ(\bw)
\label{def: delta Z} \\
\delta \bA(\hat{\bw}_t, \bw_t) & \equiv \bA(\hat{\bw}_t) - \bA(\bw_t)
\label{def: delta A} \\
\delta \bd(\hat{\bw}, \bw) & \equiv \bd(\hat{\bw}) - \bd(\bw)
\label{def: delta d} \\
\delta \bc(\hat{\bw}_t, \bw_t) & \equiv \bc(\hat{\bw}_t) - \bc(\bw_t)
\label{def: delta c}
\end{align}
Using Eqs. (\ref{def: delta Z}) through (\ref{def: delta d}), \eqref{def: x biased} can written as:
\begin{align}
\hat{\bx}_\mathrm{b} & = \bP_\tau \big[ \bZ^\rT(\bw) + \delta \bZ^\rT(\hat{\bw}, \bw)  \big]
 \bS  [\by - \bd(\bw) - \delta \bd(\hat{\bw}, \bw) ]   \nonumber \\
 & = \hat{\bx}_\ri + \bP_\tau  \delta \bZ^\rT(\hat{\bw}, \bw) \bS  [\by - \bd(\bw) ] \nonumber \\
 & \: \: \: \: -  \bP_\tau \big[ \bZ^\rT(\bw) \bS  \delta \bd(\hat{\bw}, \bw)  
 +   \delta \bZ^\rT(\hat{\bw}, \bw) \bS  \delta \bd(\hat{\bw}, \bw) \big] \,
\label{eq: x_b with error}
\end{align}
which makes use of \eqref{def: x ideal}, but replacing $\bP(\bw)$ with  $\bP_\tau$.
Note the appearance of the ideal estimate $\hat{\bx}_\mathrm{i}$ as the first term in \eqref{eq: x_b with error}.
The next step is take the expectation of both sides of \eqref{eq: x_b with error} with respect to the statistics of $\bnu$, following the same procedure that resulted in \eqref{eq: mean ideal regularized}.
This is simple, since it amounts to replacing $\hat{\bx}_\ri $ with and $\left< \hat{\bx}_\ri \right>_{\bnu}$ and $\by$ with $[ \bZ(\bw)\bx + \bd(\bw)]$ in \eqref{eq: x_b with error}. 
The result is:
\begin{align}
\left< \hat{\bx}_\mathrm{b} \right>_{\bnu} & =   \left< \hat{\bx}_\ri \right>_{\bnu}
+ \bP_\tau  \delta \bZ^\rT(\hat{\bw}, \bw) \bS \bZ(\bw) \bx   \nonumber   \\
 & \: \:  -  \bP_\tau \big[ \bZ^\rT(\bw) \bS  \delta \bd(\hat{\bw}, \bw)  
 +   \delta \bZ^\rT(\hat{\bw}, \bw) \bS  \delta \bd(\hat{\bw}, \bw) \big] \, ,
\label{eq: mean x sn -2}  \\
& =   \left< \hat{\bx}_\ri \right>_{\bnu}
+  \bG(\hat{\bw}, \bw) \bx  + \bg_1(\hat{\bw}, \bw) + \bg_2(\hat{\bw}, \bw) \, ,
\label{eq: mean x sn -1}
\end{align}
which shows that $\hat{\bx}_\mathrm{b}$ is biased relative to the ideal estimate $\hat{\bx}_\ri$.
This bias has a term $\bG(\hat{\bw}, \bw) \bx$ that is linear in $\bx$ as well as two additive terms, $\bg_1(\hat{\bw},\bw)$ and $\bg_2(\hat{\bw},\bw)$.
The $N \times N$ matrix $\bG(\hat{\bw}, \bw)$ is given by:
\begin{align}
\bG(\hat{\bw}, \bw) & \equiv \bP_\tau  \delta \bZ^\rT(\hat{\bw}, \bw) \bS \bZ(\bw)   \nonumber \\
& = \bP_\tau  \sum_{t=0}^{T-1} \delta \bA^\rT(\hat{\bw}_t, \bw_t) \bS_t \bA(\bw_t)
\nonumber \\
& = T \bP_\tau  \left<  \delta \, \bA^\rT(\hat{\bw}_t, \bw_t) \bS_t \bA(\bw_t) \right>_\tau \nonumber \\
& \equiv \bG_\tau \, ,
\label{def: G}
\end{align}
which defines the matrix $\bG_\tau$.
Eq. (\ref{def: G}) shows that, like $\bQ_\tau$, $\bG(\hat{\bw}, \bw) = \bG_\tau$ is a time average.
However, unlike $\bQ_\tau$, which can (hopefully) be approximated with knowledge of only the wavefront statistics, $\bG_\tau$ depends on the statistics of the wavefronts $\bw$ and their measurements $\hat{\bw}$ jointly.
This joint dependence is also exhibited by the $N \times 1$ vectors $\bg_1(\hat{\bw}, \bw)$ and $\bg_2(\hat{\bw}, \bw)$:
\begin{align}
\bg_1(\hat{\bw}, \bw) & \equiv - \bP_\tau \,  \bZ^\rT(\bw) \bS  \delta \bd(\hat{\bw}, \bw)   \nonumber \\
& = - \bP_\tau 
\sum_{t=0}^{T-1} \bA^\rT(\bw_t) \bS_t \delta \bc(\hat{\bw}_t,\bw_t) \nonumber \\
& = - T \bP_\tau \left< \bA^\rT(\bw_t) \bS_t \delta \bc(\hat{\bw}_t,\bw_t) \right>_\tau
\nonumber \\
& \equiv  \bg_{1\tau} \, ,
\label{def: g1} 
\end{align}
and 
\begin{align}
\bg_2(\hat{\bw}, \bw) & \equiv - \bP_\tau \, \delta  \bZ^\rT(\bw) \bS  \delta \bd(\hat{\bw}, \bw)    \nonumber \\
& = - \bP_\tau 
\sum_{t=0}^{T-1} \delta \bA^\rT(\bw_t) \bS_t \delta \bc(\hat{\bw}_t,\bw_t) \nonumber \\
& = - T \bP_\tau \left< \delta \bA^\rT(\bw_t) \bS_t \delta \bc(\hat{\bw}_t,\bw_t) \right>_\tau
\nonumber \\
& \equiv   \bg_{2\tau} \, ,
\label{def: g2} 
\end{align}
which define the vectors $\bg_{1\tau}$ and $\bg_{2\tau}$.
Then, using Eqs.~(\ref{def: G}) through (\ref{def: g2}), the expression for the mean of the biased estimate in \eqref{eq: mean x sn -1} becomes:
\begin{equation}
\left< \hat{\bx}_\mathrm{b} \right>_{\bnu}  = 
\left< \hat{\bx}_\mathrm{i} \right>_{\bnu} + \bG_\tau  \bx  + \bg_{1\tau} + \bg_{2 \tau} \, ,
\label{eq: mean x sn}
\end{equation}
which is expressed in terms of the (acceptable) bias of the ideal estimator given in \eqref{eq: approx mean ideal regularized}.

The next order of business is to calculate the covariance of the biased estimator.
Using \eqref{def: x biased} and the definition of the noise covariance $\bC$, several lines of algebra demonstrate:
\begin{equation}
< \hat{\bx}_\mathrm{b} \hat{\bx}_\mathrm{b}^\rT >_{\bnu}  - < \hat{\bx}_\mathrm{b} >_{\bnu} < \hat{\bx}_\mathrm{b}^\rT >_{\bnu}
= \bP_\tau \bH(\hat{\bw})  \bP_\tau  \, 
\label{eq: var x biased}
\end{equation}
which is similar in appearance to covariance of the ideal estimator in \eqref{eq: ideal cov H}, and where the $N \times N$ matrix $\bH(\hat{\bw})$ is defined as [see also \eqref{eq: H def}]:
\begin{align}
\bH(\hat{\bw}) & \equiv \bZ^\rT(\hat{\bw}) \bS \bC \bS \bZ(\hat{\bw}) \nonumber \\
& = \sum_{t=0}^{T}  \bA^\rT(\hat{\bw}_t) \bS_t \bC_t \bS_t \bA(\hat{\bw}_t) \, .
\label{def: H}
\end{align}
Here, one should note that $\bH(\hat{\bw})$ can be calculated with only the measured wavefronts.

\subsection{Bias-Corrected Estimation}

The final estimator we present is the \emph{bias-corrected estimator}, and it is culmination of our efforts. 
If the various assumptions are met, the bias-corrected estimator converges to the ideal estimator, which we are taking to be acceptably biased, as per the remarks in Sec.~\ref{sec: regression}\ref{sec: ideal}.
The bias-corrected estimator, $\hat{\bx}_\mathrm{c}$, is defined as:
\begin{equation}
\hat{\bx}_\mathrm{c} \equiv \bP_\tau \bQ_\tau 
\big( \bG_\tau + \bP_\tau \bQ_\tau \big)^{-1}
\big( \hat{\bx}_\mathrm{b} - \bg_{1\tau} - \bg_{2\tau}  \big) \, ,
\label{eq: x c}
\end{equation}
which is expressed in terms of $\hat{\bx}_\mathrm{b}$, the biased estimator from \eqref{def: x biased}. 
With the help of \eqref{eq: approx mean ideal regularized} and \eqref{eq: mean x sn}, is trivial to verify that:
\begin{equation}
\left< \hat{\bx}_\mathrm{c} \right>_{\bnu}  
= \bP_\tau \bQ_\tau \bx
= \left< \hat{\bx}_\ri \right>_{\bnu} \, ,
\label{eq: mean csn}
\end{equation}
thus demonstrating that the bias-corrected estimator has the same (acceptable) bias that the ideal estimator does.
Using Eqs. (\ref{eq: var x biased}) and (\ref{eq: x c}) , one immediately sees that the covariance matrix of the bias-corrected estimator is:
\begin{align}
 < \hat{\bx}_\mathrm{c} \hat{\bx}_\mathrm{c}^\rT >_{\bnu} &  - < \hat{\bx}_\mathrm{c} >_{\bnu} < \hat{\bx}_\mathrm{c}^\rT >_{\bnu} \,
=   \bP_\tau \bQ_\tau \big( \bG_\tau + \bP_\tau \bQ_\tau \big)^{-1} \nonumber \\
& \: \times \: \:   \bP_\tau \bH(\hat{\bw})\bP_\tau  \, 
 \big( \bG_\tau + \bP_\tau \bQ_\tau \big)^{-\rT}  \bQ_\tau  \bP_\tau \, ,
\label{eq: var x c} \\
\overset{\beta = 0}{=}  &   \big( \bG_\tau + \mathbb{1} \big)^{-1} 
\bP_\tau \bH(\hat{\bw})\bP_\tau \big( \bG_\tau + \mathbb{1} \big)^{-\rT}
\label{eq: var x c no reg}
\end{align}
where $^{-\rT}$ indicates the transpose of the inverse of a matrix (unlike $\bQ_\tau$ and $\bP_\tau$, the matrix $\bG_\tau$ is not symmetric), and $\mathbb{1}$ is the identity matrix (in this case, $N \times N$).
As shown in \eqref{eq: var x c no reg}, when no regularization is applied (so, $\beta = 0$ and $\bP_\tau = \bQ_\tau^{-1}$), the expression simplifies considerably. 
It is important to notice that in the limit of no wavefront error (i.e., $\hat{\bw} = \bw$), $\bG_\tau = 0$, and the covariance of the bias-corrected estimator is exactly that of the ideal estimator shown in \eqref{eq: ideal cov H}, i.e., $ \bP_\tau \bH(\bw)\bP_\tau $.
In other words, the effect of the wavefront measurement error is manifested in the $\bG_\tau$ matrix, which serves to inflate the covariance of the bias-corrected estimator.

\subsection{Monte Carlo Proxies}

In order to calculate the bias-corrected estimate and its error covariance matrix, we need computable proxies for the "$_\tau$ \nolinebreak quantities": $\bQ_\tau, \;\bP_\tau,  \; \bG_\tau, \; \bg_{1\tau} $ and $\bg_{2\tau}$.
For this, we will rely on the Monte Carlo sampling capabilities discussed in Sec.~\ref{sec: Wavefronts}.

Let us assume that the Monte Carlo engine provides $L$ samples of the AO residuals in the set $\{ \breve{\bw}_l \}$.  
If $L$ and $T$ are large enough we have:
\begin{align}
\bQ_\mathrm{mc} & \equiv  \frac{T}{L}\sum_{l=0}^{L-1} \bA^\rT(\breve{\bw}_l) \bS_t \bA(\breve{\bw}_l) \,  
\label{def: Qmc}  \\
& \approx  T \left< \bA^\rT(\breve{\bw}_l) \bS_t \bA(\breve{\bw}_l) \right>_\rE 
\label{eq: Q ensemble mean} \\
& \approx  T \left< \bA^\rT(\bw_t) \bS_t \bA(\bw_t) \right>_\tau 
 = \bQ_\tau  \, , 
\label{eq: Q temporal mean}
\end{align}
where the weight matrix $\bS_t$ is taken to be a constant that does not depend on $t$ (as in the Part I simulations) [see \eqref{eq: S_t blocks}].
\footnote{Using the same matrix $\bC_t$ for all values of $t$ prevents minimum mean-squared error estimation in the presence of noise that depends on $t$ (such as shot-noise), which requires $\bS_t = \bC_t^{-1}$.
}

Recalling \eqref{eq: P is a time average}, the matrix $\bP_\mathrm{mc}$ is a Monte Carlo proxy for the Kalman gain, we define:
\begin{equation}
%\bP_\mathrm{mc} & \equiv \mathrm{RegularizedInverse}  \left[ \beta \bXi +  \bQ_\mathrm{mc}\, ; \, \gamma  \right] \label{def: Pmc} 
\bP_\mathrm{mc} \equiv \left( \beta \bXi +  \bQ_\mathrm{mc} \right)^{-1} \, .
\label{def: Pmc}
\end{equation}
Notice that when $\beta > 0$ the inversion in \eqref{def: Pmc} is regularized, which mitigates Monte Carlo sampling errors.
($\beta$ was zero in the Part I simulations.)
It is interesting to note that the regularization term, $\beta \bXi $, introduced in \eqref{eq: ideal cost} to mitigate  the effects of ill-conditioning in the ideal estimator, could prove beneficial for the Monte Carlo approximations as well.
Future studies may consider other forms of regularized inversion in place of \eqref{def: Pmc}.
Statisticians have created a large literature on estimating inverse covariance matrices from sample covariance matrices, which will not summarized here.
However, the authors did find Ref.~[\citenum{WierPeeters_InvMonteCarloCov_CSDA16}], in which its authors provided several regularized estimators of the inverse convariance matrix given a sample covariance matrix, to be potentially useful.
The most important thing to remember is that the inverse of sample covariance matrix converges to the true inverse covariance matrix in the large sample limit.

The Monte Carlo approximations of the matrix $\bG_\tau$ and the vectors $\bg_{1\tau}$ and $\bg_{2\tau}$ rely not only on the ability to generate Monte Carlo wavefronts, but on the capacity to simulate the WFS measurements as well.
Starting with \eqref{def: G}, the matrix $\bG_\tau$ admits the following Monte Carlo approximation:
\begin{align}
\bG_\tau & \equiv T \bP_\tau  \left<  \delta \bA^\rT(\hat{\bw}_t, \bw_t) \bS_t \bA(\bw_t) \right>_\tau \nonumber \\
& \approx T \bP_\tau  \left<  \delta \bA^\rT(\hat{\breve{\bw}}_l, \breve{\bw}_l) \bS_t \bA(\breve{\bw}_l) \right>_\rE \nonumber \\
& \approx \frac{T}{L} \bP_\mathrm{mc} \sum_{l=0}^{L-1}  \delta \bA^\rT(\hat{\breve{\bw}}_l, \breve{\bw}_l) \bS_t \bA(\breve{\bw}_l) \nonumber \\
& \equiv \bG_\mathrm{mc}
\label{def: Gmc}
\end{align} 
where $\hat{\breve{\bw}}_l$\ is a simulated measurement of a Monte Carlo wavefront $\breve{\bw}_l$.
Similarly, the vectors  $\bg_{1\tau}$ and $\bg_{2\tau}$ in Eqs.~(\ref{def: g1}) and (\ref{def: g2}) have the Monte Carlo approximations:
\begin{align}
\bg_{1\mathrm{mc}} & \equiv -  \frac{T}{L} \bP_\mathrm{mc}
\sum_{l=0}^{L-1} \bA^\rT(\breve{\bw}_l) \bS_t \delta \bc(\hat{\breve{\bw}}_l,\breve{\bw}_l) \; , \: \mathrm{and}
\label{def: g1mc}  \\
\bg_{2\mathrm{mc}} & \equiv -  \frac{T}{L} \bP_\mathrm{mc}
\sum_{l=0}^{L-1} \delta \bA^\rT(\breve{\bw}_l) \bS_t \delta \bc(\hat{\breve{\bw}}_l,\breve{\bw}_l) \, .
\label{def: g2mc}
\end{align}
One useful attribute of approximating the various $_\tau$ quantities with Monte Carlo methods is that all of the nonlinear dependencies on the wavefront $ \bw_t $ are treated without approximation.

\subsection{Numerical Implementation of Regression Algorithm}

Armed with various Monte Carlo approximations, we are now in a position to calculate a Monte Carlo version of the bias-corrected estimate, $\hat{\bx}_\mathrm{cmc}$, based on Eqs.~(\ref{def: x biased}) and (\ref{eq: x c}).
Simply put, the idea is to replace all of the $_\tau$ quantities with their "$_\mathrm{mc}$" counterparts.
Thus, the Monte Carlo bias-corrected estimate is given by:
\begin{align}
\hat{\bx}_\mathrm{cmc} \equiv & \: \bP_\mathrm{mc} \bQ_\mathrm{mc} 
\big(  \bG_\mathrm{mc} +  \bP_\mathrm{mc} \bQ_\mathrm{mc} \big)^{-1}  \; \times
\nonumber \\
& \big\{ 
 \bP_\mathrm{mc}   \bZ^\rT(\hat{\bw}) \bS  [\by - \bd(\hat{\bw}) ]
- \bg_{1\mathrm{mc}} - \bg_{2\mathrm{mc}}  \big\} \, .
\label{eq: x cmc}
\end{align}
Note that the realizable version of the bias corrected estimate, is in fact the "Monte Carlo bias-corrected estimate," $\hat{\bx}_\mathrm{cmc}$.
Similarly, the covariance of this estimate comes from \eqref{eq: var x c}, again by replacing the $_\tau$ quantities with their Monte Carlo counterparts.

Below, we summarize the computations required to implement the regression algorithm.  
All of the steps below have minimal memory requirements, and are suited for handling millions of exposures and for GPU-based computations.
The algorithm's steps are as follows:
\begin{enumerate}
\item{Use Monte Carlo sampling to calculate: $\bQ_\mathrm{mc}, \; \bP_\mathrm{mc}, \; \bG_\mathrm{mc} , \; \bg_\mathrm{1mc}$ and $\bg_\mathrm{2mc} \,$.}

\item{As the science camera intensities $\{\by_t \}$ and wavefronts $\{ \hat{\bw}_t \}$ are observed, accumulate the sums:
  \begin{itemize}
	\item{$\bZ^\rT(\hat{\bw}) \bS  [\by - \bd(\hat{\bw}) ]  = \sum_{t=0}^{T-1} \bA^\rT(\hat{\bw}_t) \bS_t [\by_t - \bc(\hat{\bw}_t)]$.}

	\item{ $\bH(\hat{\bw}) =
 	\sum_{t=0}^{T}  \bA^\rT(\hat{\bw}_t) \bS_t \hat{\bC}_t \bS_t \bA(\hat{\bw}_t) \, $.  
	Here, we approximate $\bC_t$, which is the noise covariance matrix of $\bnu_t$, with $\hat{\bC}_t$, which approximates unknown true intensity (in photon units) with the measured number of photon counts when calculating the variance of the shot-noise contribution to the noise covariance of the science camera measurements.
	Approximating the true photon count rate with the measured photon count rate is acceptable since the latter is an unbiased estimator of the former.\cite{StatisticalOptics}
	Note that $\bH(\hat{\bw})$ is only used to calculate the error covariance matrix of the estimates, not the estimates themselves, so inaccuracies are less critical.}
\end{itemize}
}

\item{ Calculate Monte Carlo bias-corrected estimator from \eqref{eq: x cmc}.}

\item{ Calculate the Monte Carlo approximation to the error covariance matrix of $\hat{\bx}_\mathrm{c}$ [see \eqref{eq: var x c}]: 
$\bP_\mathrm{mc} \bQ_\mathrm{mc} \big( \bG_\mathrm{mc} + \bP_\mathrm{mc} \bQ_\mathrm{mc} \big)^{-1}
   \bP_\mathrm{mc} \bH(\hat{\bw})\bP_\mathrm{mc} 
 \big( \bG_\mathrm{mc} + \bP_\mathrm{mc} \bQ_\mathrm{mc} \big)^{-\rT}  \bQ_\mathrm{mc}  \bP_\mathrm{mc}$}

\item{If the nonlinearity in the NCPA coefficients needs to be treated (this nonlinearity does effect the estimate of the all components of $\bx$), re-linearize about the estimate, as per \eqref{eq: total Taylor}, and start  over.}

\end{enumerate}

\subsubsection*{Mitigating Roundoff Error}

The above algorithm for the Monte Carlo bias-corrected estimator is designed to accommodate millions of millisecond images and Monte Carlo samples.
The loops in which the needed sums are accumulated must carry enough floating point precision so that the roundoff error does not significantly impact the accuracy of the value contributed by a given increment after millions of increments have already been totaled.
For example, in calculating $\bQ_\mathrm{mc}$ from \eqref{def: Qmc}, the $l$\underline{th} increment is the matrix $\bA^\rT (\breve{\bw}_l) \bS_t \bA(\breve{\bw}_l)$. 
The reader will recall that $\bA$ is horizontally partitioned into an NCPA portion and an exoplanet portion [see \eqref{def: x A}], which can have different scalings.
Accurately accumulating matrices with scalings that vary by orders-of-magnitude over millions of samples can be problematic.  
One tactic is to find a diagonal matrix $\bM$ such that $\bM \bA^\rT (0) \bS_t \bA(0)\bM$ has columns with a maximum absolute value of unity [$\bA(0)$ indicates that functions in $\bA$ are evaluated for a flat wavefront, and $\bS_t$ is assumed to be fixed)].
Once $\bM$ is specified, the sum $\bQ'_\mathrm{mc}$ is accumulated increments of $ \bM\bA^\rT (\breve{\bw}_l) \bS_t \bA(\breve{\bw}_l) \bM$.

Note that $\bA(\breve{\bw}_l) \bx = \bA(\breve{\bw}_l) \bM \bM^{-1} \bx $,  and replacing  $\bA(\breve{\bw}_l)$ with $\bA(\breve{\bw}_l) \bM$  corresponds to a new coordinate system in which $\bx' = \bM^{-1} \bx$.
This transformation law also serves to transform the estimate, so $\hat{\bx} = \bM \hat{\bx}'$.

\section{Convergence Issues}\label{sec: convergence}

The bias-corrected estimator relies on Monte Carlo approximations of the $_\tau$ quantities: $\bQ_\tau, \;\bP_\tau,  \; \bG_\tau, \; \bg_{1\tau} $ and $\bg_{2\tau}$.
The assumptions underlying the validity of these approximations are essentially the same, so we will focus our discussion on the approximation of $\bQ_\tau$ with $\bQ_\mathrm{mc}$, as shown in 
Eqs.~(\ref{def: Qmc}) though~(\ref{eq: Q temporal mean}).

The definition of $\bQ_\mathrm{mc}$ is analogous to $\bQ(\bw_t)$ as defined in \eqref{def: Q} [see also \eqref{def: c Z}].
The Monte Carlo engine draws the $L$ samples $\breve{\bw}_0, \dots, \,  \breve{\bw}_{L-1} $ from some distribution function (most likely a multivariate normal).
The elements of the matrix  $ \bA^\rT(\breve{\bw}_l) \bS_t \bA(\breve{\bw}_l) $ are nonlinear functions of the vector $ \breve{\bw}_l $, but it is clear that as $L$ increases, the sum in \eqref{def: Qmc} approaches the mean of  $ \bA^\rT(\breve{\bw}_l) \bS_t \bA(\breve{\bw}_l) $  with respect to the probability distribution from which the samples $\{ \breve{\bw}_l \}$ are drawn.
For a multivariate normal distribution and linear functions $\bA$, the statistical properties of $\bQ_\mathrm{mc}$ corresponding to a finite samples size $L$ are quite well understood, as they are given by the \emph{Wishart distribution}.
Further, the inverse of a matrix drawn from the Wishart distribution is governed by the \emph{inverse Wishart distribution}.\cite{MatrixMultivariateDistributions} 
For nonlinear functions $\bA$, as we have, results are hard to obtain.
However, modern parallel processing, particularly on GPUs, make generating large numbers of samples practical, particularly for sampling from a multivariate normal, so we will assume that $L$ is large enough for \eqref{eq: Q ensemble mean} to be valid.
In the Part I simulations of the bias-corrected estimate, the Monte Carlo wavefronts were sampled from a multivariate normal distribution with the same 2\underline{nd} order statistics (i.e., mean and covariance matrix) as the true wavefronts, and there were $L = 480,000$ sample wavefronts.

Whether or not \eqref{eq: Q temporal mean} is valid is a more subtle matter.
It requires that the ensemble mean of the matrix  $ \bA^\rT(\breve{\bw}_l) \bS_t \bA(\breve{\bw}_l) $ is close to the temporal mean of the random process  $ \bA^\rT(\bw_t) \bS_t \bA(\bw_t) $ and that the finite time interval, $T$, is of sufficient duration for
$ \frac{1}{T} \sum_0^{T-1} \, \bA^\rT(\bw_t) \bS_t \bA(\bw_t) $ to approach its temporal mean.
This is most easily understood when the wavefronts are produced by a stationary random process, so the statistics can be expressed as infinite time averages.
In practice, the best one hopes for is a quasi-stationary random process, in which the statistical properties vary over time-scales that much longer than the time-scale of correlations between values of the variables in $\bw_t$. 
In the Part I simulations, the temporal correlation time-scale of the AO residuals was roughly $10^{-2}$ s, while the observation time, $T$, was 240~s.

The assumption that the Monte Carlo engine draws from the distribution governing the wavefronts $\{ \bw_t \}$ is an assumption that is still more difficult to realize.
It was in fact, the violation of this assumption that limited the accuracy of the bias-corrected estimates given in Part I.
In those simulations, the wavefronts generated by the AO simulator were not consistent with multivariate normality, even when we considered only short time intervals consisting of 100 milliseconds, let alone the entire 240~s.  
We know this because the time-series obtained from several pupil pixels (randomly chosen) failed the Anderson-Darling test for normality.
\footnote{The Anderson-Darling test is for consistency with univariate normality that is robust to correlation in the data.} 
Since multivariate normal deviates must satisfy univariate normality on the individual variables, this showed that the wavefront vectors were not multivariate normal.

The source of the non-normality of AO residual wavefronts in the Part I simulations is probably the AO system, as opposed the atmospheric phase screens themselves.
To the extent that the atmospheric optical path difference (OPD) is the sum of the OPDs caused by density (and water vapor content, depending on the wavelength) fluctuations the individual air elements in the column of atmosphere, it is generally thought that the central limit theorem should apply, leading to normally distributed OPD.\cite{Roddier81, StatisticalOptics}
In the Part I simulations, the atmospheric OPD is summation of six statistically independent OPDs, each corresponding to a layer of turbulence.
However, the intensity measured by camera in WFS is necessarily nonlinear in the OPD, and the DM will have a response to the DM command that is at least somewhat nonlinear.
(In the Part I simulations, the DM surface height was modeled as a 2D cubic spline interpolator.)
Thus, the AO system applies nonlinear filtering to a normal random process, giving the AO residual non-normal character.

\subsection{Validation Via Simulation}\label{sec: validation}

\begin{figure}
\begin{tabular}[t]{c}
	\includegraphics[width=.99\linewidth,clip=]{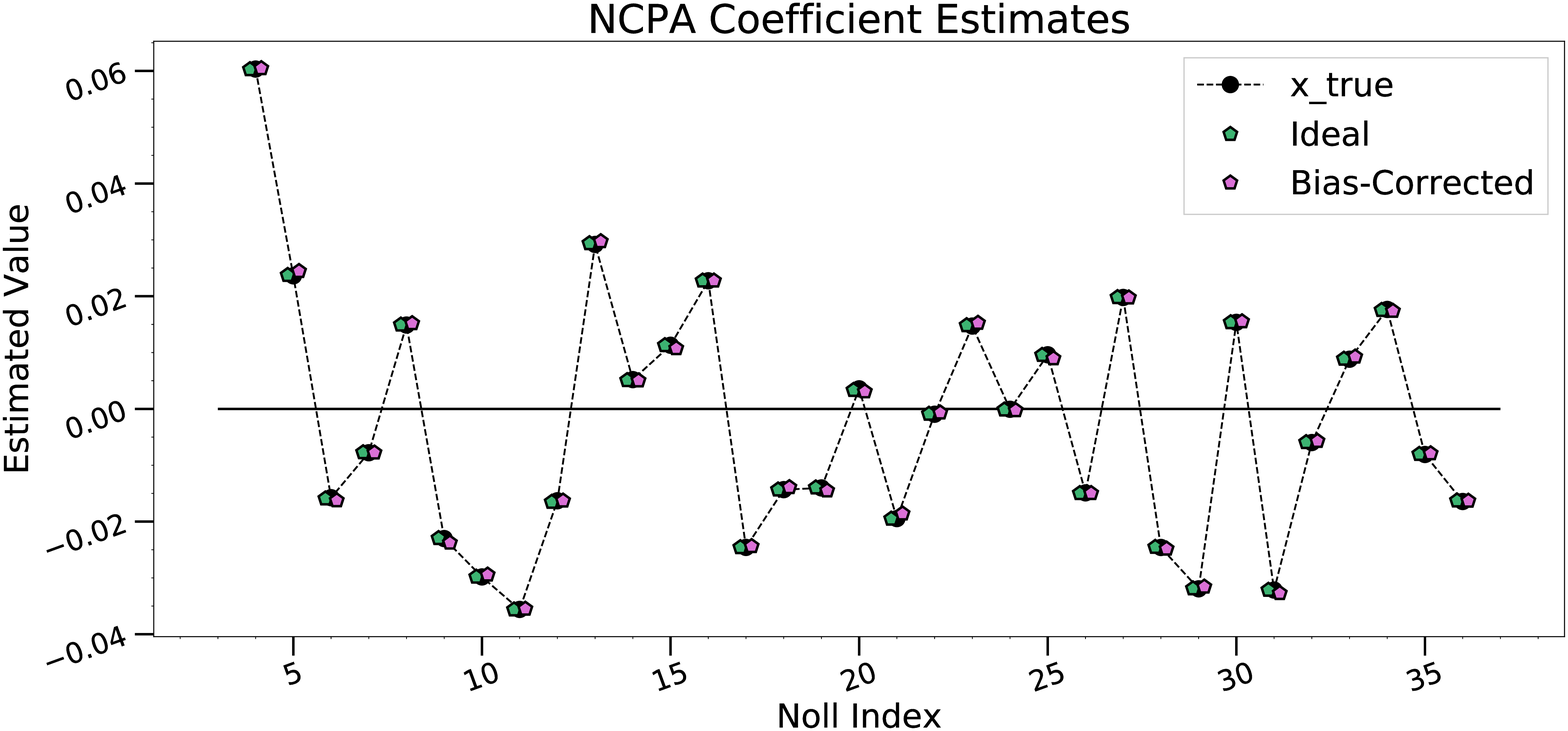} \\
	\includegraphics[width=.99\linewidth,clip=]{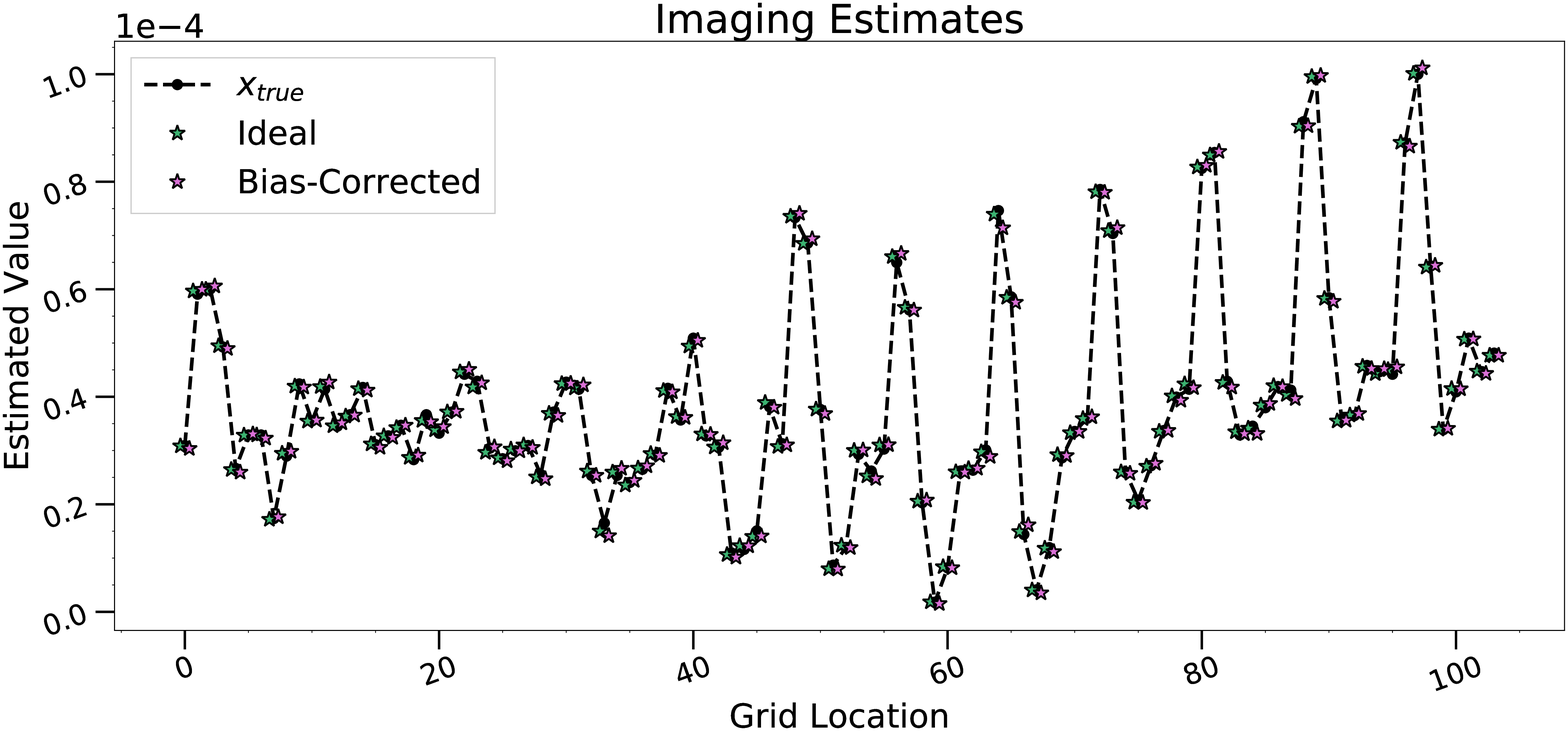}
\end{tabular}
	\caption{\small Bias-corrected and na\"ive estimates for the experiment described in Sec.~\ref{sec: convergence}\ref{sec: validation}, in which white noise was used for the AO residual phases.  The na\"ive estimates would not fit within the vertical range of the plot.
	The experimental setup corresponds to the Phase B simulation in the Part I article (apart from replacing the AO residuals with white noise).
	\emph{upper panel:}  NCPA coefficient values and estimates.  \emph{lower panel:}  Image coefficient estimates.
	The error bars are too small to see in the figure.  See text for more detail.}
	\label{fig: white noise results}
	\vspace{-4mm}
\end{figure}

Part I presented a simulation study of the ideal, na\"ive and bias-corrected estimators along with extensive discussion and analysis.
However, the fact that the AO residual phase values were not consistent with a multivariate normal distribution prevented the assumptions underlying the bias-corrected from being fully satisfied.
Here, we present another simulation in which we removed the problem of non-normality in the AO residuals as a validation exercise.
In this simulation, the $P=1976$ AO residual pixel values (at a given time-step $t$) were replaced by white noise with a standard deviation of $ 0.6 \,$radian, corresponding to a Strehl ratio of about $0.7$.
The white noise phase screens were drawn independently from each other, so there was no temporal correlation in the AO residuals. 
In this way, it was simple to draw both the Monte Carlo and AO residual wavefronts from same distribution.
This simulation had $T = 10^6$ time-steps and used $L = 10^6$ Monte Carlo samples.
Other than these changes, the simulation set-up was exactly as in the Phase B simulation of Part I,
so, there was a $13 \times 8$ grid of exoplanet image points, 33 NCPA coefficients (corresponding to Zernike polynomials, without piston tip or tilt), and the star's magnitude was 8.
The same models of the pyramid WFS and Lyot stellar coronagraph were used.

The ideal and bias-corrected estimation results shown in Fig.~\ref{fig: white noise results}.
The na\"ive estimates did not fit within the plotting range for many of the points.
The error bars (which come from diagonal elements of the error covariance matrix of the estimate), are too small to be seen in the figures.
In simulation studies such as this, we can judge accuracy of the error covariance matrix, $\bC_{\hat{\bx}}$, of an (unbiased) estimator with the following $\chi^2$ measure of data misfit:
$\chi^2 = \frac{1}{N} (\bx - \hat{\bx})^\rT \bC_{\hat{\bx}}^{-1}(\bx - \hat{\bx})$.
With $\bchi^2$ defined this way, it has an expected value of unity, but biases in the estimate will make the value of $\chi^2$ larger, often by orders-of-magnitude.
For estimates shown in Fig.~\ref{fig: white noise results}, the $\chi^2$ value for the ideal estimate was 1.1, and for the bias-corrected estimate it was 4.5.
For comparison, the corresponding $\chi^2$ value for the na\"ive estimate was $7.7 \times10^4$.

\section{Conclusions}

Whatever one's criticisms of the techniques presented here, the stubborn reality of ground-based direct imaging of exoplanets is that the intensity measured by the science camera is a nonlinear function of the telescope's optical system with hardware aberrations, the kHz AO residual wavefront, and the putative exoplanets.
For example, if one attempts to measure the NCPA with a Zernike wavefront sensor, the 2\underline{nd} order statistics of the AO residual wavefronts must be included in the estimation,\cite{Vigan_ZELDA_NCPA_AA19} and errors in the assumed statistics and the coronagraph model will limit the achievable contrast.

Several studies have established that the success of an exoplanet survey mission depends heavily on being able to achieve high contrast within a sky-angle of several $\lambda/D$ of the host stars.\cite{Stark_ExoEarthYield14, Brown_PlanetSearch15}
However, the region of the image plane within several $\lambda/D$ is where the current post-processing techniques are least effective: Angular differential imaging (ADI), which relies on the Earth's field rotation, is compromised by the short arclength traversed by the planet, and spectral differential imaging (SDI) is not particularly helpful because the stellar point-spread function (PSF) does not stretch much with increasing wavelength close to the center.
Furthermore, all of the differential imaging methods suffer from self-subtraction issues that are problematic for point-sources and that worsen with extended sources.\cite{Marois_SOSIE, Rameau_ADI_SDI_limits15, Mawet_SmallNumStatsSpeckle14} 
In contrast, the regression method presented here does not rely on differential imaging and suffers from no self-subtraction artifacts.
Additionally, the field rotation that underpins ADI can be included in this regression with little difficulty, as can spectroscopic capabilities in the coronagraph optical train.

It is now becoming increasingly clear that by resolving the time-scale over which the AO residual wavefronts evolve, millisecond imaging provides a wealth of information that is otherwise unavailable, and a new generation of noiseless and near-noiseless detection technologies are becoming available.\cite{Saphira_eAPD14,Mazin_MKIDS18}
One of the key features of millisecond imaging is the fact that the intensity measured by the science camera resolves the temporal modulation of the PSF by the AO residual.
This modulation of the PSF results in differences in temporal behavior of the planetary and stellar speckles, including "dark speckle" instants in time in which the planet's contrast is enhanced.\cite{Labeyrie_DarkSpeckle95}
The WFS telemetry is obviously quite complimentary to the millisecond science camera images, since it can provide an estimate of wavefront propagating through the coronagraph.
So far, the regression methods presented here are the only methods proposed so far that utilize the WFS telemetry, hardware models, and millisecond imaging to make a joint estimate of the NCPA and the exoplanet image.
The bias-corrected estimator presented here is but a modest first attempt to treat errors in wavefront estimates, which create biases that limit the achievable contrast.

Future success in ground-based direct imaging of exoplanets will require accurate models of the WFS and coronagraph optical trains along with error budgets.
The problem of jointly estimating the free parameters that specify the hardware models and the NCPA can be approached iteratively in an expectation-maximizatin (EM) framework.
Use of the EM algorithm to correct for modeling errors in electric field conjugation (for space-based missions) has already been highly successful in laboratory settings, working up to $~10\%$ wavelength bandwidth.\cite{SunHeEFC18}
But, even more important and challenging will be advances in the measurement of the AO residual wavefronts and quantifying their statistical properties, including chromatic and amplitude (i.e., scintillation) effects.\cite{Devaney_ChromaticAO,Jolissaint_ChromaticAO}
The need to include the statistical properties of the AO residual wavefronts has long been recognized in the literature on PSF reconstruction.\cite{Perrin_PSFstructure_ApJ03, Wagner_PSFrecon_JATIS18,Schiller_COFFEE19,Fetick_PSF19}

In the authors' view, the direct imaging of exoplanets has much in common with a completely different observational astronomy problem: the detection of gravitation waves  with the LIGO experiment.
The precision underlying the success of LIGO has depended on removing systematic effects where possible, modeling them when not possible, and state-of-the art statistical analysis.\cite{LIGO2017}
One lesson that could be used from LIGO is building systems and designing them so they are amenable to modeling.
For example, LIGO's heavy mirrors that make the Fabry-Perot cavity are suspended by fibers make out of pure silicon crystals.
The choice of pure silicon was made so that the vibrational modes would be known \emph{a-priori}.
Such choices in an exoplanet imaging system could include field stops in focal planes to limit high spatial-frequency content, metrology systems to monitor alignments, and so-on.  
It is in this spirit of looking toward the future challenges that this regression method has been presented.

\section*{Acknowledgments}
The authors would like to thank: the anonymous referees, P. Scott Carney, Jared Males, Olivier Guyon, He Sun, and Len Stefanski.
This work has been supported by the NSF (Award \#1600138) and the Heising-Simons Foundation (Grant \#2020-1826).

\section*{Disclosures}
The authors declare no conflicts of interest.

%%\bibliographystyle{osajnl.bst} %don't include this line
%\bibliography{exop}

\end{document}